\documentclass[conference]{IEEEtran}
\usepackage[numbers]{natbib}
\usepackage{amsmath,amssymb,amsfonts,bm}
\usepackage{algorithmic}
\usepackage{graphicx}
\usepackage{textcomp}
\usepackage{xcolor}
\usepackage{hyperref}
\usepackage{booktabs}
\usepackage{subcaption}
\usepackage{float}
\usepackage{multirow}
\def\BibTeX{{\rm B\kern-.05em{\sc i\kern-.025em b}\kern-.08em
    T\kern-.1667em\lower.7ex\hbox{E}\kern-.125emX}}
\begin{document}

\title{Modeling Wallet-Level Behavioral Shifts Post-FTX Collapse: An XAI-Driven GLM Study on Ethereum Transactions \\}

\author{
\IEEEauthorblockN{Benjamin Gillen}
\IEEEauthorblockA{\textit{Mathematics} \\
\textit{USC}\\
Los Angeles, USA \\
bgillen@usc.edu}
\and
\IEEEauthorblockN{Rashmi Ranjan Bhuyan}
\IEEEauthorblockA{\textit{Data Science and Operations} \\
\textit{USC}\\
Los Angeles, USA \\
bhuyanr@usc.edu}
\and 
\IEEEauthorblockN{Gourab Mukherjee}
\IEEEauthorblockA{\textit{Data Science and Operations} \\
\textit{USC}\\
Los Angeles, USA \\
gourab@usc.edu}
\and
\IEEEauthorblockN{Austin Pollok}
\IEEEauthorblockA{\textit{Data Science and Operations} \\
\textit{Finance and Business Economics} \\
\textit{USC}\\
Los Angeles, USA \\
pollok@usc.edu}
}

\maketitle

\begin{abstract}
The Ethereum blockchain plays a central role in the broader cryptocurrency ecosystem, enabling a wide range of financial activity through the use of smart contracts. This paper investigates how individual Ethereum wallets responded to the collapse of FTX, one of the largest centralized cryptocurrency exchanges. Moving beyond price-based event studies, we adopt a bottom-up approach using granular wallet-level data. We construct a representative sample of Ethereum addresses and analyze their transaction behavior before and after the collapse using an explainable artificial intelligence (XAI) framework. Our proposed framework addresses data scarcity in high-resolution wallet-level daily transactions by employing a calibrated zero-inflated generalized linear fixed effects model. Our analysis quantifies distinct shifts in transaction intensity and stablecoin usage, highlighting a flight to safety within the ecosystem. These findings underscore the value of a bottom-up methodology for quantifying the user-level impact of blockchain-based shocks, offering insights beyond traditional price-level analysis through wallet-level data.

\end{abstract}


\section{Introduction}
Blockchain offers a decentralized alternative to traditional data systems, enabling secure, transparent, and tamper-resistant transactions. Its adoption across sectors, from finance to healthcare, reflects its potential to streamline operations and reshape digital infrastructure. As regulatory clarity improves and markets for stablecoins and tokenized assets expand, blockchain’s ability to challenge conventional business models continues to attract institutional interest and capital investment \cite{coinbase_state_of_crypto_2025}.

Bitcoin, the first blockchain, is primarily limited to serving as a cryptocurrency. In contrast, Ethereum has emerged as the leading public blockchain for smart contracts, enabling transactional capabilities through programmable accounts, or wallets. These Ethereum wallets, represented by addresses, store account balances, transaction times, and executable code. As transaction volumes grow and more economic activity moves on-chain, wallet-level analysis becomes essential for tracking flows \cite{cucuringu_clustering_uniswap_traders}, valuing assets, and auditing stablecoins to ensure they meet their mandates, such as maintaining a peg to the U.S. dollar.

As adoption of on-chain and off-chain applications continues to increase, the broader blockchain ecosystem experiences persistent volatility and periodic shocks. Understanding how individual wallets respond to these disruptions provides insight into the behavioral dynamics of crypto markets, which are often obscured by aggregate indicators like BTC or ETH price movements. Analyzing wallet-level activity reveals important heterogeneity in user behavior but also presents a challenge. Daily data for many wallets are sparse, with most addresses inactive on a typical day. To address this, we use zero-inflated models that explicitly account for the high frequency of zeros. We adopt generalized linear fixed effects models, which accommodate repeated observations and allow partial pooling across users. This makes them well-suited for blockchain data that are both sparse and heterogeneous \cite{bolker2009glmm}. Unlike purely predictive machine learning methods, our framework yields interpretable parameter estimates and supports explainable AI by identifying which covariates and temporal effects influence wallet-level behavior.

A major shock occurred in 2022, when a wave of credit contagion culminated in the collapse of FTX, at the time the third-largest cryptocurrency exchange. Allegations of fraud, customer fund mismanagement, and rapid bankruptcy exposed vulnerabilities in centralized exchange infrastructure. Ethereum, central to the DeFi ecosystem, was particularly affected. The crisis intensified when it was revealed that Alameda Research, a trading firm closely tied to FTX, held substantial assets in FTT, FTX’s native token. Binance’s withdrawal from a proposed acquisition triggered mass withdrawals, leading to a liquidity crunch and eventual collapse \cite{FTX_downfall}. The involvement of high-profile entities and real-time information flow through social media amplified volatility and investor panic. Despite the severity of the disruption, the shock remained largely contained within the crypto ecosystem and had limited spillover into traditional financial markets.

To date, few studies have examined how individual wallets respond to systemic shocks. Existing research has largely focused on price-based measures. For example, \cite{crypto_volume_volatility, crypto_volatility} document negative price reactions and increased volatility following the collapse. Reference \cite{bayesian_structural_counterfactual} uses Bayesian structural models to show that Ethereum’s price declined below counterfactual estimates. These studies underscore the internal contagion dynamics of crypto markets and suggest that their connections to traditional finance remain limited.

In this paper, we take a bottom-up, data-driven approach to assess the impact of the FTX collapse on the Ethereum ecosystem by studying wallet-level behavior. Rather than relying on price movements formed through off-chain speculation, we focus on user-level responses observable on-chain. Using transaction-level data, we estimate heterogeneous effects across wallets with zero-inflated generalized linear models that account for individual variation and macroeconomic conditions. By modeling wallet features such as transaction frequency, token flows, and stablecoin usage before and after the event, we uncover structural changes in user behavior.

Our analysis reveals distinct patterns in wallet activity between April 2022 and January 2023. As shown in Fig.~\ref{fig:transaction} and Fig.~\ref{fig:eth_price}, the share of active wallets rose from around six percent in April to thirteen percent by mid-October. Following the FTX collapse, this share fell to nine percent and continued declining through year-end. Average transactions per wallet followed a similar trend, peaking in late September, falling by twenty percent after the collapse, and then gradually returning to earlier levels. These movements broadly track Ethereum-USD price trends and motivate the inclusion of asset prices and macro indicators in our models.

This study makes three primary contributions. First, we show that individual wallets respond heterogeneously to systemic shocks. Aggregate price metrics obscure this variation, which we recover through exploratory analysis and formal estimates, such as those shown in Fig.~\ref{fig:wallet_mean}, ~\ref{fig:wallet_zero}. Second, we quantify wallet-level responses using zero-inflated fixed effects models with wallet-specific intercepts and time-varying splines. This allows us to capture sparsity in wallet activity and identify both average and user-specific shifts. Third, we provide uncertainty quantification through bootstrapped confidence intervals, allowing for assessment of robustness and statistical significance.

Our modeling approach emphasizes interpretability and behavioral insight, aligning with the goals of explainable AI. Rather than optimizing for prediction, we focus on identifying structural drivers of behavior across user types. These findings are relevant to researchers, developers, and policymakers who seek to understand the dynamics of blockchain ecosystems during periods of market stress.

The remainder of this paper is organized as follows. Section II describes the data and sampling process. Section III presents the statistical models used to estimate wallet-level effects. Section IV reports the results. Section V concludes with a discussion of key findings and future directions.

\section{Data}

We analyze Ethereum wallet-level transaction activity using a combination of on-chain and off-chain data spanning April 1, 2022, to January, 1 2023. The dataset construction proceeds in two stages first wallet address selection and then transaction retrieval.

\subsection{Ethereum Wallet-Level Transaction Data}

We leverage data from the XBlock-ETH project \cite{xblock_eth}, a curated dataset that extracts and chronologically organizes raw Ethereum blockchain transactions into a structured, analysis-ready format. It provides comprehensive coverage of all on-chain activity, including sender and receiver addresses, timestamps, gas usage, and token transfer information, which facilitates wallet-level analysis.

We download the XBlock-ETH block transaction files corresponding to our study period. To determine exactly which files we need, we use Etherscan’s  \texttt{getblocknobytime} API method to identify the blocks mined at our interval’s start and end timestamps. This lets us pinpoint the block range -- and thus the specific block transaction files -- corresponding to our time range. Next, we initialize an empty set. For each block transaction file, we read the information line by line. Each line contains information on a single transaction, including the block number and sender address. This sender address corresponds to the wallet -- technically an externally owned account (EOA) -- that initiated the transaction. We add each sender address to our set if two conditions are met: first, the transaction's block falls within our specified block range, and second, the sender address is not already contained within our set. After processing all lines across all block transaction files, we obtain a set of all EOAs that initiated at least one transaction within our time range. From this set, we uniformly sample 50,000 addresses for deeper analysis.

For each wallet in the sample, we queried the API for Alchemy, a node infrastructure provider, using the \texttt{alchemy\_getAssetTransfers}  method, which returns all standard transactions either initiated by or received by a particular address. Transactions are categorized by token or protocol type, including:

\begin{itemize}
\item \textbf{External} – standard ETH transfers initiated by externally owned accounts (EOAs),
\item \textbf{Internal} – ETH transfers triggered by smart contract execution,
\item \textbf{ERC-20} – fungible token transfers conforming to the ERC-20 standard,
\item \textbf{ERC-721} – non-fungible token (NFT) transfers under the ERC-721 standard,
\item \textbf{ERC-1155} – hybrid transfers supporting both fungible and non-fungible tokens.
\end{itemize}

Each transaction record includes metadata such as the block number, transaction hash, sender and receiver addresses, asset type, transferred value, and transaction category. Using this data, we can record the total units of Ethereum and stablecoin that a wallet buys and sells over time. We classify ERC-20 transactions as stablecoin transactions if the token traded was one of the seven principal stablecoins that existed in 2022: USD Coin, Tether, Binance USD, TrueUSD, Gemini Dollar, Dai, and Frax.



\subsection{Explanatory Covariates}
We include two exogenous time series as covariates in our model, each designed to capture an economically meaningful channel through which wallet-level transaction behavior may respond to broader market forces:

\begin{enumerate}
\item \textbf{Ethereum price} (\texttt{ETHPRICE}).

Wallet activity is closely tied to the price of Ethereum through both mechanical and behavioral channels. On the mechanical side, wallets seeking to extract or deploy a fixed dollar amount will transact fewer tokens when the price is high and more when the price is low. On the behavioral side, rising prices often trigger increased trading activity driven by momentum effects or fear of missing out (FOMO), while falling prices may induce hesitation or loss-aversion behavior. Including \texttt{ETHPRICE} helps capture both these mechanisms—how price levels influence transaction sizes, and how price trends may affect the decision to trade at all.
\item \textbf{Six-month U.S. Treasury yield} (\texttt{RF6M}).

The six-month Treasury rate serves as a proxy for the short-term risk-free return available in traditional financial markets. As this rate increases, the relative attractiveness of holding volatile crypto assets like Ethereum may decline, prompting some wallets—particularly those engaged in yield-seeking or treasury-type functions—to reduce exposure. Conversely, a lower risk-free rate may encourage reallocation into riskier digital assets. Including \texttt{RF6M} allows us to separate crypto-specific transaction patterns from those driven by broader macro-financial conditions, especially shifts in opportunity cost of capital. We use the six-month maturity rather than the one-month Treasury to smooth over short-term volatility while remaining responsive to macroeconomic changes.
\end{enumerate}


\section{Models and Methodology}

\subsection{Model of Wallets' Responses to FTX}








From our sample of 50,000 wallets, we identify the subset of wallets that made at least five transactions during the observation window. Let \( m \) denote the number of wallets in this subset. We track the transaction activity of these \( m \) wallets on a daily basis, beginning from April 1, 2022, over a period of \( n \) days, indexed by \( t = 1, \ldots, n \). For the remaining wallets (i.e., those with fewer than five transactions), we monitor only aggregate daily summary statistics.

For each of the $m$ wallets we track four daily activity streams, $E_{it}^{b}$ and $E_{it}^{s}$ record the the total units of Ethereum that wallet $i$ buys and sells at time $t$, while $S_{it}^{b}$ and $S_{it}^{s}$ denotes their corresponding quantity bought and sold of stablecoin. We model each of these responses with a zero-inflated generalized linear model  with covariates  $\bm{X}_{t}$. We write the specifications of the model for a generic response $Y_{it}$ and obtain the four separate models, with distinct coefficients per model, by substituting $Y_{it} \in \{E_{it}^{b}, E_{it}^{s}, S_{it}^{b}, S_{it}^{s}\}$.

\[
Y_{it}\;=\;
\begin{cases}
0, & \text{with probability }\pi_{it},\\[6pt]
\texttt{Gamma}\!\bigl(k,\mu_{it}/k\bigr), & \text{with probability }1-\pi_{it},
\end{cases}
\]
where the zero-inflated probability $\pi_{it}$ is modeled with a logit link and a log link is used to model conditional mean , $\mu_{it} = \mathbb{E}[Y_{it}\mid Y_{it}>0]$. 
\begin{align}
\log \mu_{it}
      &= {\alpha_i}
       + f(t)
       + \bm\zeta^{\!\top}\bm{X}_{t}, \label{eq:glm_link_1}  \\
\operatorname{logit}\,\pi_{it}
      &= {\gamma_i}
       + g(t)
       + \bm\kappa^{\!\top}\bm{X}_{t}\label{eq:glm_link_2}, 
\end{align}

with, \(\alpha_i,\bm\zeta\) and \(\beta_i,\bm\kappa\), being the wallet specific intercepts and slopes with respect to our chosen auxiliary information. The shape parameter for the Gamma distribution is common across all wallets. The functions, $f(t)$ and $g(t)$ represent the temporal variation through smooth functions. Both are modeled as natural cubic splines with \(10\) effective degrees of freedom. Concretely, for any time \(t\) we write
\[
f(t)=\sum_{k=1}^{10}\beta_k\,B_k(t),\qquad
g(t)=\sum_{k=1}^{10}\delta_k\,B_k(t),
\]
where \(B_1(t),\dots,B_{10}(t)\) form a B-spline basis where each basis function is a piece wise cubic polynomial. Since the basis is fixed, \(\boldsymbol\beta=(\beta_1,\dots,\beta_{10})^{\!\top}\) and \(\boldsymbol\delta=(\delta_1,\dots,\delta_{10})^{\!\top}\) enter the model linearly.

\subsection{Model Estimation}
The key task is to quantify the coefficients \(\alpha_i,\gamma_i,\bm\zeta,\bm\kappa\), the spline functions $f(t)$, $g(t)$ through $\bm \beta, \bm\gamma$, as well as the
common shape parameter \(k\) that drive the
zero-inflated Gamma model. We also define \(\mathbf I_{it}^0:=\mathbf I\{Y_{it}=0\}\) and \(\mathbf I_{it}^+=1-\mathbf I_{it}^0\), the variables denoting whether the response was zero or not. The contribution of $(i,t)$ to the log likelihood, $\ell_{it}$ is
\begin{align*}
\ell_{it}
   =  \mathbf I_{it}^+\Bigl\{&\,
          \log(1-\pi_{it})
        + k\log\!\bigl(k/\mu_{it}\bigr)
        - \log\Gamma(k)\\
        &+ (k-1)\log Y_{it}
        - k\,Y_{it}/\mu_{it}\Bigr\} + \mathbf I_{it}^0\,\log\pi_{it}
        \,,
\end{align*}
where, and $\pi_{it}$, $\mu_{it}$ are described in \eqref{eq:glm_link_1}-\eqref{eq:glm_link_2}. Summing over all wallets and days yields the total likelihood $\ell(\Phi)=\sum_{i=1}^{m}\sum_{t=1}^{T}\ell_{it}$ where $\Phi$ describes the set of all the parameters involved as
\(\Phi=\bigl(
      \{\alpha_i,\gamma_i\}_{i = 1}^n,k,\boldsymbol\beta,\boldsymbol\delta,\bm\zeta, \bm\kappa
     \bigr).\)
     
Maximizing the closed-form log-likelihood \(\ell(\Phi)\) reduces to a standard optimization problem. We obtain the resulting maximum-likelihood estimates with the \texttt{glmmTMB} package \citep{mcgillycuddy2025parsimoniously}. We also recover the fitted splines curves by simple matrix products \( \hat f(t)=\mathbf B(t)^{\top}\hat{\boldsymbol\beta}\) and \( \hat g(t)=\mathbf B(t)^{\top}\hat{\boldsymbol\delta}\).

\subsection{Confidence Interval}

The model estimation method implicitly provides us standard errors on the parameters $\Phi$. This is useful for the parameters corresponding to our covariates, but to quantify the confidence interval in the smooth time functions, we need to estimate uncertainty in \(f(t)=\mathbf B(t)^{\top}\boldsymbol\beta\). While we can draw a separate interval at every time $t$, the probability that all of those intervals cover the true curve would be far below the nominal level. To alleviate this, we need to construct simultaneous bands, that cover the entire trajectory of $f(t)$ with the required probability.

We employ a parametric bootstrap method to estimate the confidence interval. First, we fit the model and obtain the estimated parameters $\hat\Phi$, and correspondingly $\hat f(t)$. Next, for each bootstrap replicate \(b=1,\dots,B\) we perform three 
simple steps:
\begin{enumerate}
    \item Simulate a new response vector \(Y_b^{\!*}\sim P_{\hat\Phi}\) from the fitted model
    \item Refit the model to \(Y_b^{\!*}\) to get a new coefficient vector \(\hat{\boldsymbol\beta}^{\!*}_b\)
    \item Evaluate the corresponding curve
\(\hat f^{\!*}_b(t)=\mathbf B(t)^{\top}\hat{\boldsymbol\beta}^{\!*}_b\) and record 
\(S^{\!*}_b=\max_t\lvert \hat f^{\!*}_b(t)-\hat f(t)\rvert\), the largest absolute deviation over time. 
\end{enumerate}
After repeating the loop \(B\) times, the \((1-\alpha)^\text{th}\) percentile of the collection \(\{S^{\!*}_b\}\) gives a critical value \(c_{\alpha}\). Our final \((1-\alpha)\%\) simultaneous confidence band is therefore \(\hat f(t)\pm c_{\alpha}\).
\section{Results}

\subsection{Descriptive Results}

\begin{figure}
    \centering
    \includegraphics[width=0.75\linewidth]{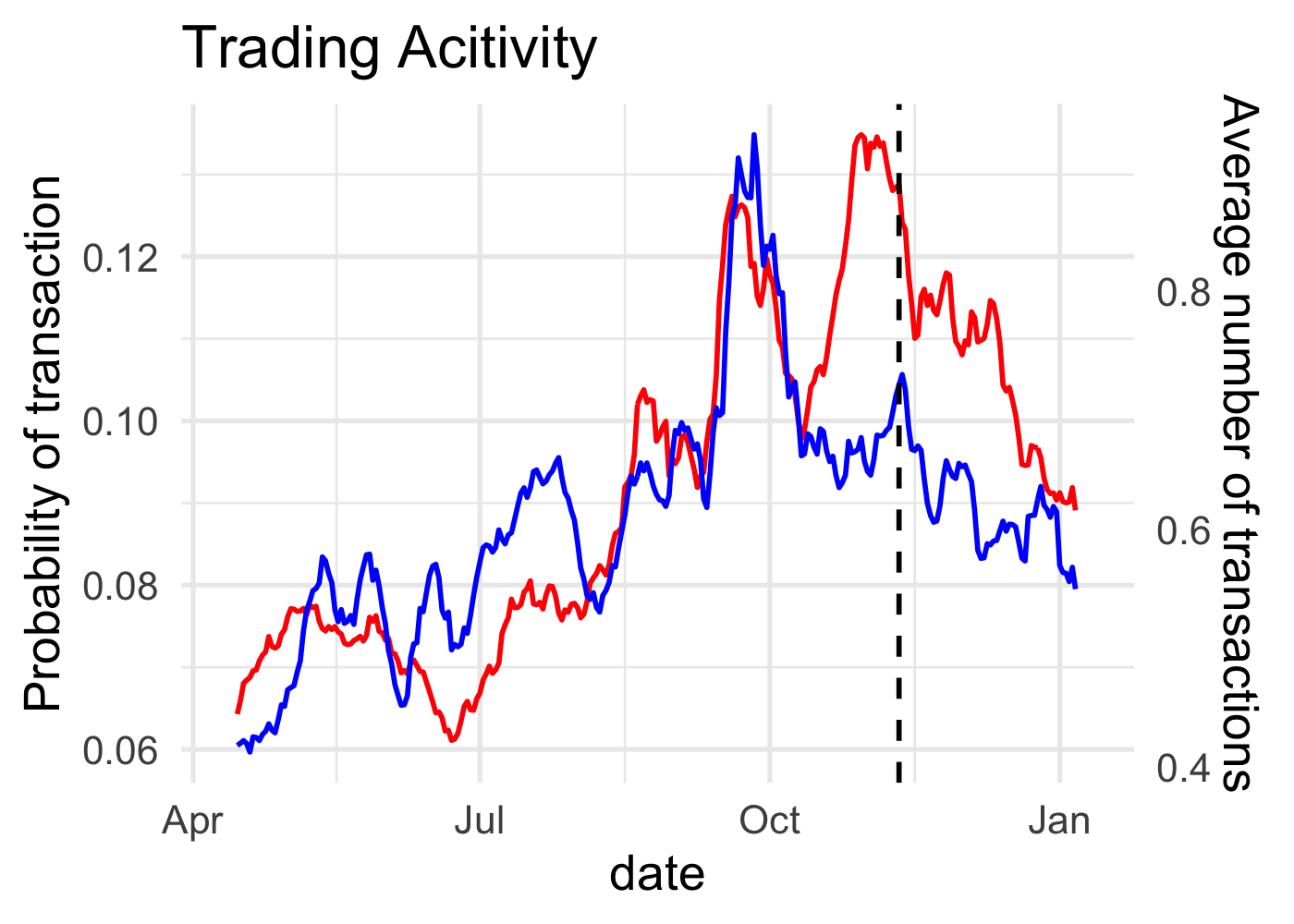}
    \caption{Trading activity across Ethereum and stablecoin across active wallets, 10-days moving average, probability of a transaction (in red), average number of transactions (in blue).}
    \label{fig:transaction}
\end{figure}

\begin{figure}
    \centering
    \includegraphics[width=0.75\linewidth]{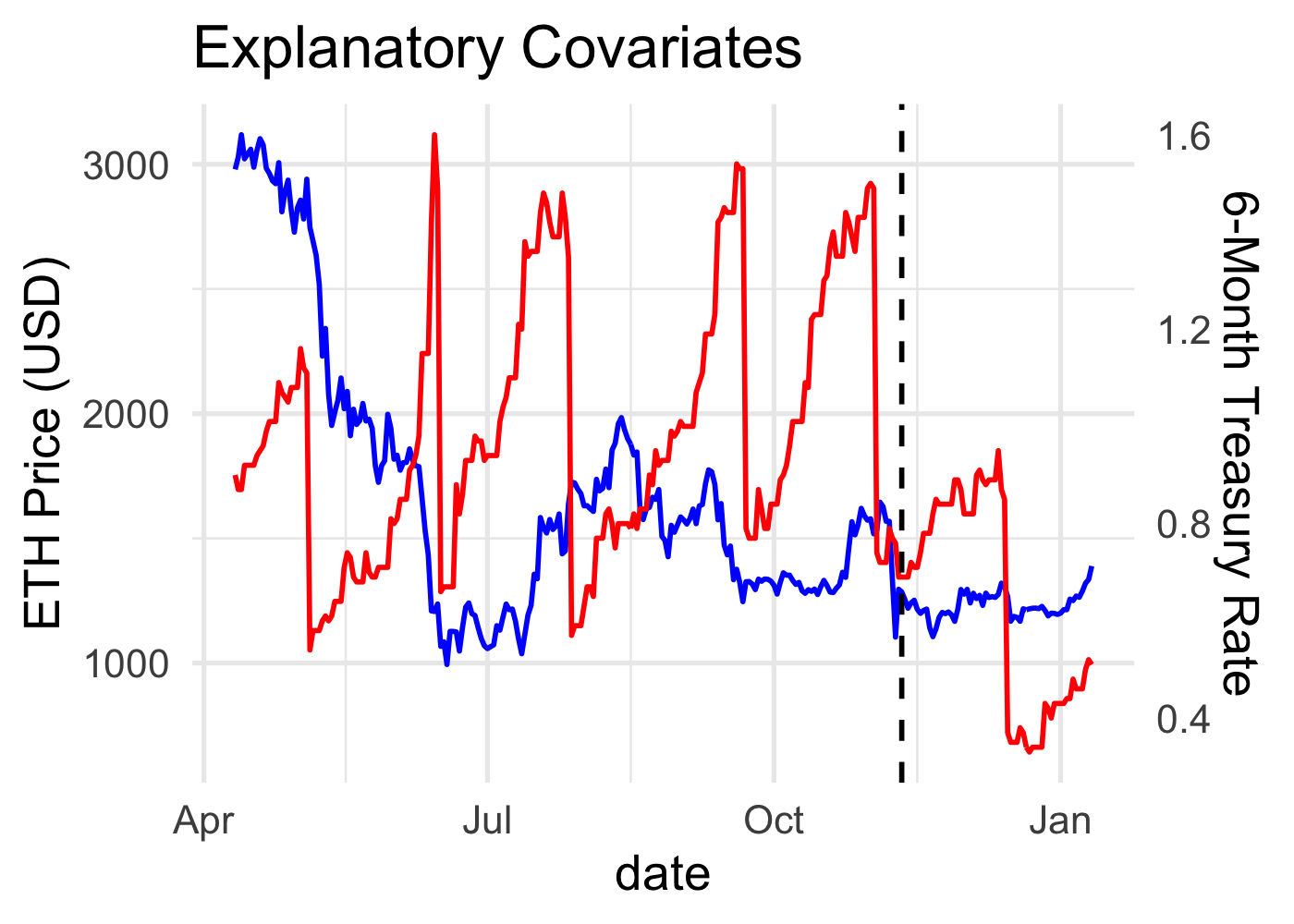}
    \caption{Explanatory covariates, Ethereum price in US Dollars (in blue), 6-month treasury rate (in red).}
    \label{fig:eth_price}
\end{figure}

Fig.~\ref{fig:transaction}-\ref{fig:eth_price} summarizes the wallet behavior and the Ethereum-USD price, as well as the risk-free rate during the period of our study. Specifically, in Fig.~\ref{fig:transaction}, we see the percentage of active wallets. The probability that a randomly chosen wallet trades at all climbs from roughly 6–7 \% in April to a peak of 13 \% by mid-October.  Immediately after the FTX insolvency that share drops to about 9 \% and drifts lower through December, indicating a consistent contraction in market participation. Fig.~\ref{fig:transaction} also shows the mean trade count. This metric rises steadily through the spring, peaks near one transaction per active wallet in late September, and then falls in two steps, first a sharp 20 \% decline in the week after FTX, followed by a less sharp slowdown towards early-summer levels. Finally, we show the Ethereum price and treasury rate fluctuations during this period in Fig.~\ref{fig:eth_price}. Comparing it with the percentage of active wallets, we see a co-movement that encourages us to use them as covariates to model wallet activity. 

\begin{figure}
\centering
\begin{subfigure}{.22\textwidth}
  \centering
  \includegraphics[width=\linewidth]{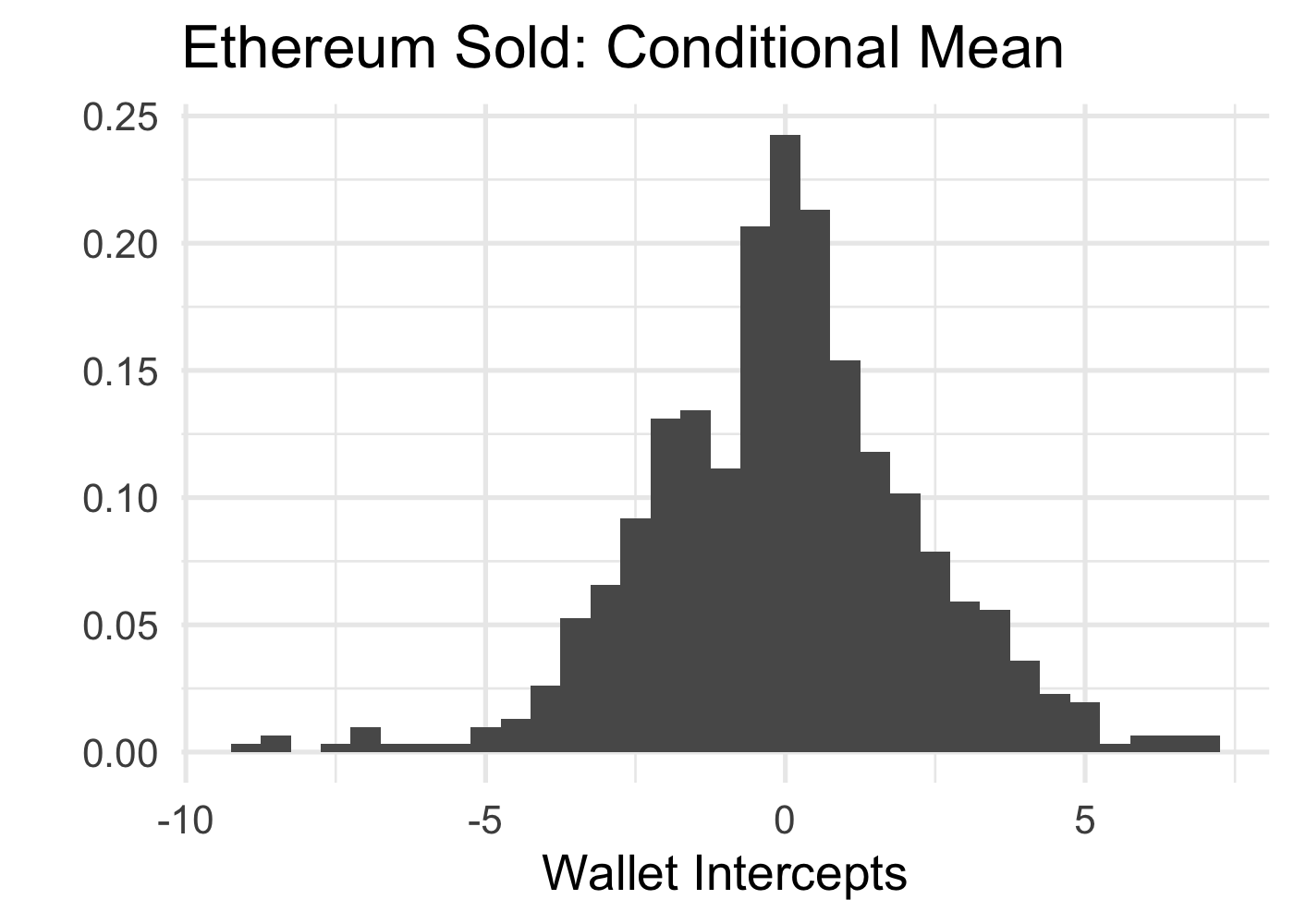}
  \caption{}
  \label{fig:wallet_mean}
\end{subfigure}%
\begin{subfigure}{.22\textwidth}
  \centering
  \includegraphics[width=\linewidth]{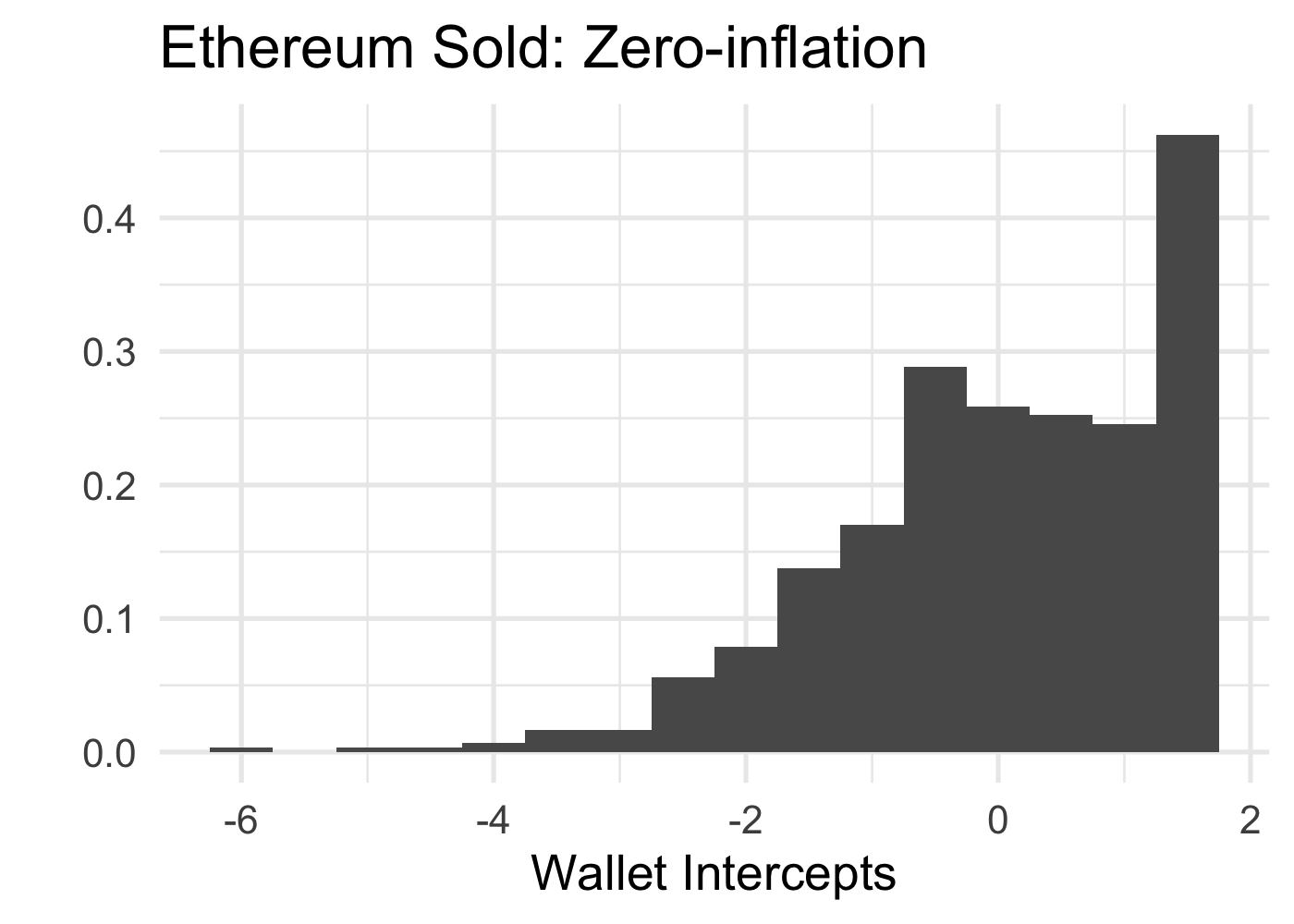}
  \caption{}
  \label{fig:wallet_zero}
\end{subfigure}
\caption{Distribution of wallet-specific intercepts \(\alpha_i\) (conditional mean) and \(\gamma_i\) (zero-inflation) from the Ethereum sale model. }
\label{fig:wallet_eth_sales}
\end{figure}

\begin{figure}[]
\centering
\begin{subfigure}{.22\textwidth}
  \centering
  \includegraphics[width=\linewidth]{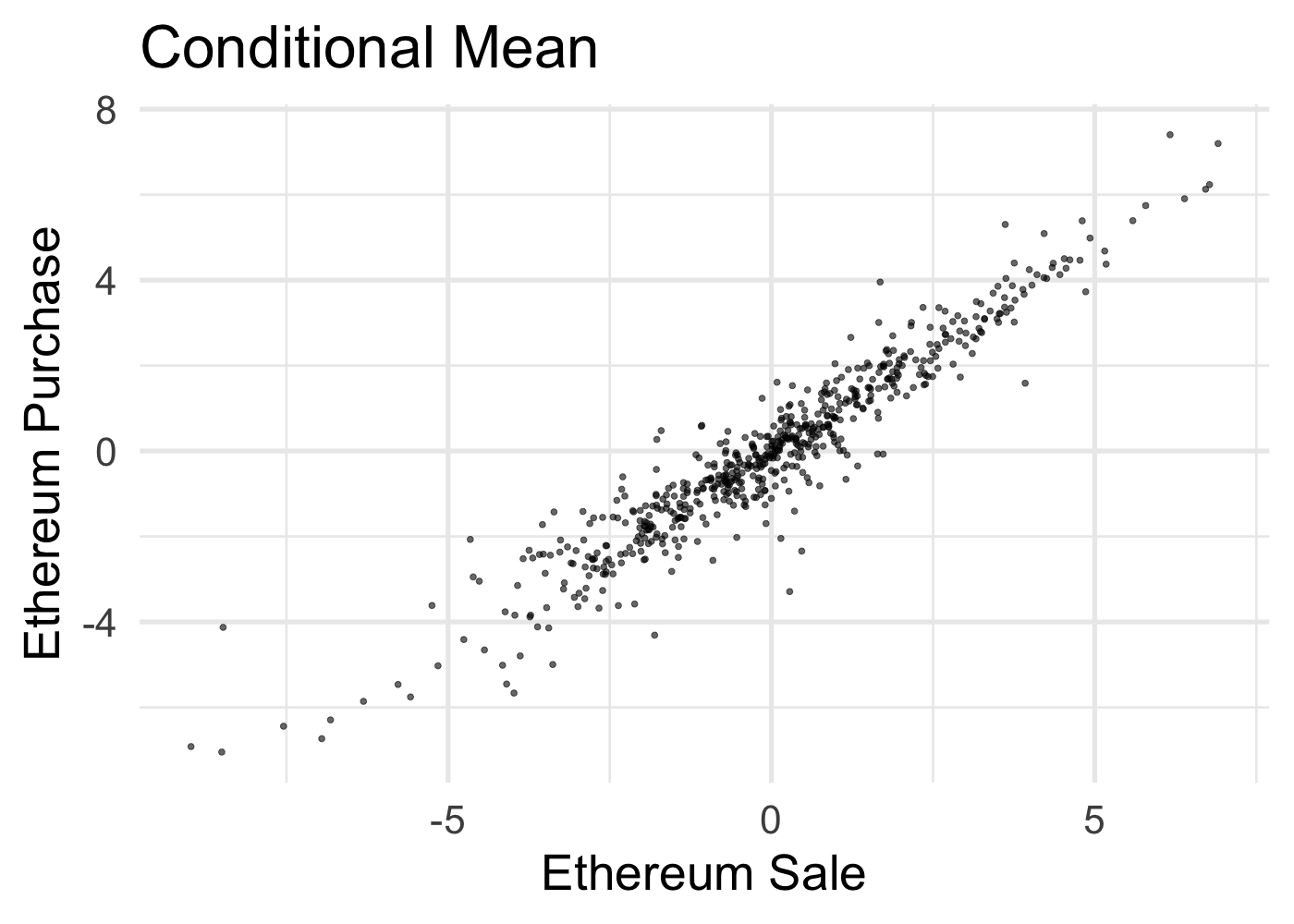}
  \caption{}
  \label{fig:scatter_ep_es}
\end{subfigure}%
\begin{subfigure}{.22\textwidth}
  \centering
  \includegraphics[width=\linewidth]{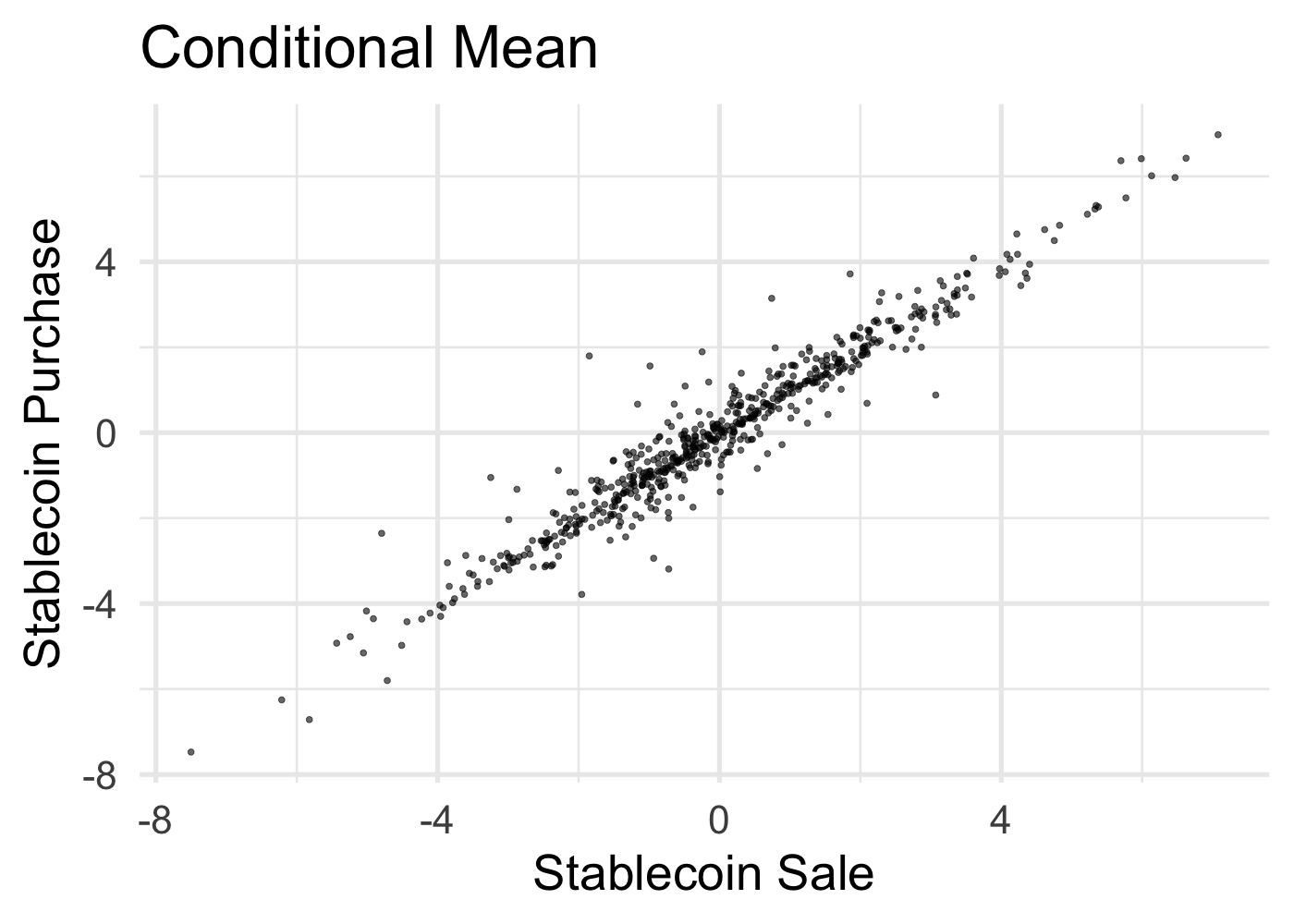}
  \caption{}
  \label{fig:scatter_sp_ss}
\end{subfigure}
\caption{Scatter plot of wallet-level intercepts from the conditional mean model, with each point representing a wallet's estimated intercepts for ETH and stablecoin transactions.}
\label{fig:scatterplots_ee_ss}
\end{figure}

\begin{figure}[]
\centering
\begin{subfigure}{.22\textwidth}
  \centering
  \includegraphics[width=\linewidth]{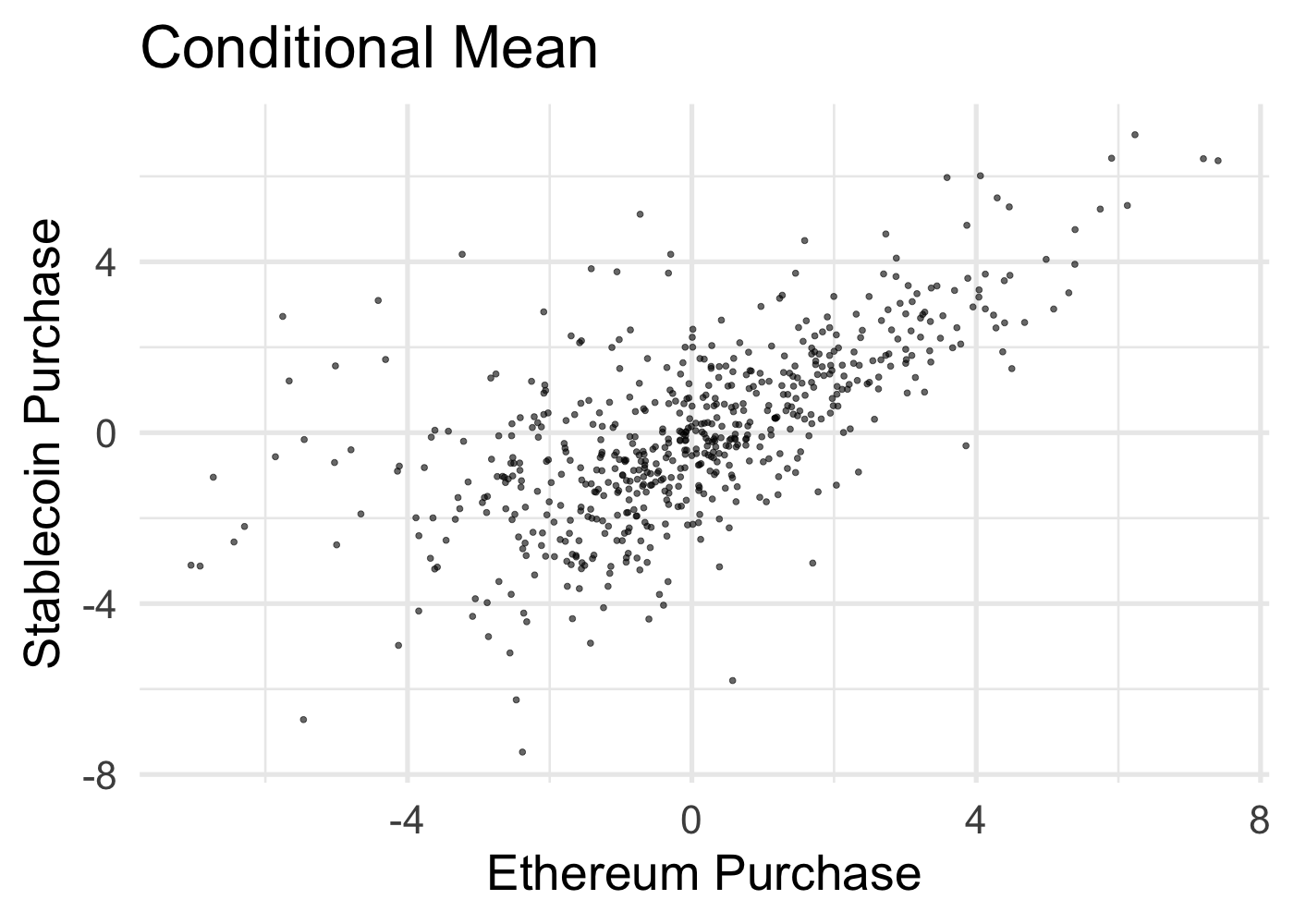}
  \caption{}
  \label{fig:scatter_sp_ep}
\end{subfigure}%
\begin{subfigure}{.22\textwidth}
  \centering
  \includegraphics[width=\linewidth]{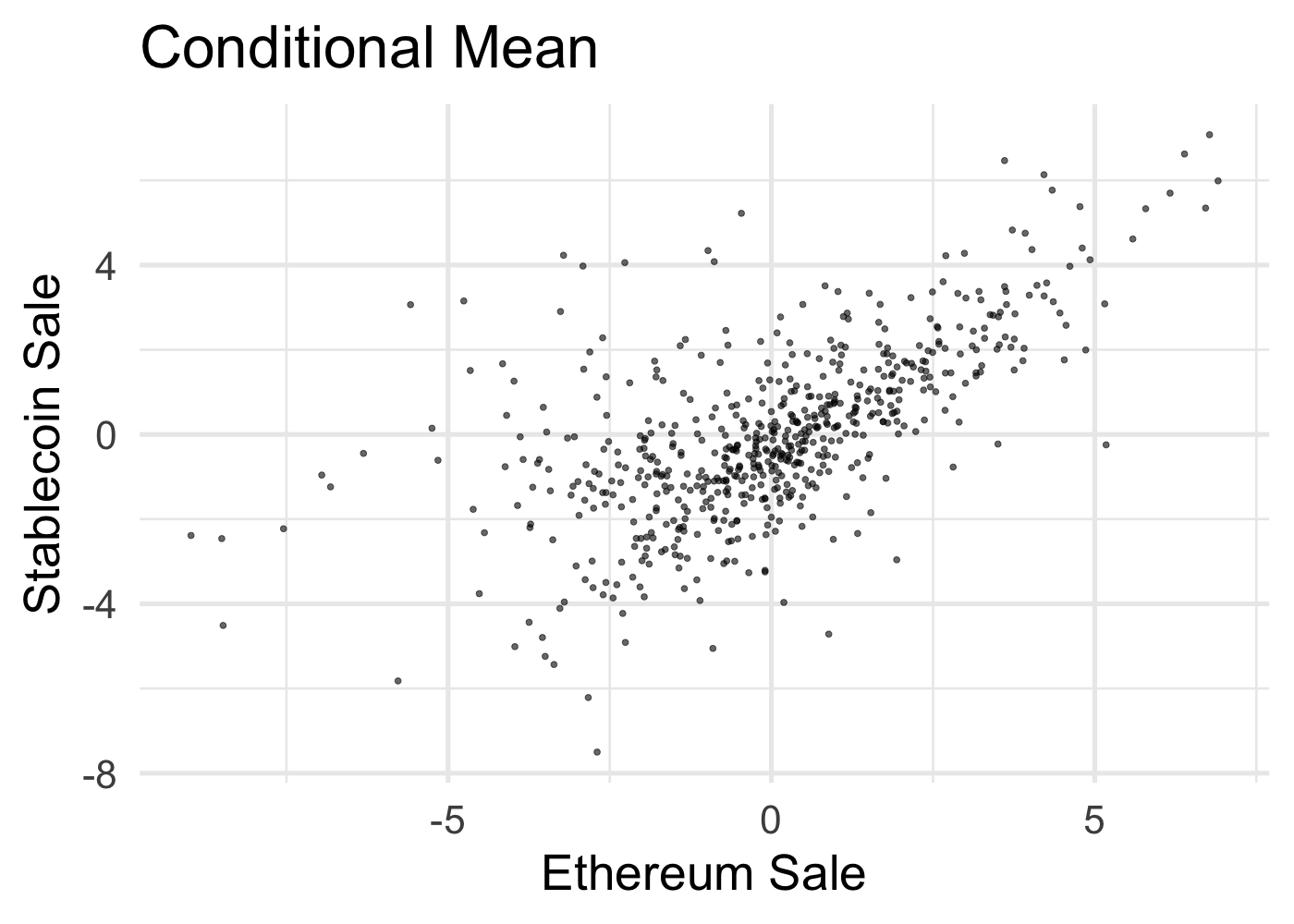}
  \caption{}
  \label{fig:scatter_ss_es}
\end{subfigure}
\caption{Wallets in the upper-left transact more ETH and less stablecoin per trade, while those in the lower-right show the opposite, based on wallet-level conditional mean intercepts.}
\label{fig:scatterplot_sep_ses}
\end{figure}

Fig.~\ref{fig:wallet_eth_sales} shows the distribution of wallet-specific intercepts for the Ethereum sales model. While conditional mean intercepts are symmetric, zero-inflation intercepts exhibit strong left-skewness. This indicates that most wallets have low trading propensity, while a small number are persistently active. These patterns are consistent across all four models, as confirmed by the skewness values reported in Table~\ref{tab:skew_wallet}. Fig. ~\ref{fig:scatterplots_ee_ss} plots the wallet-level conditional mean intercepts for Ethereum and stablecoin sales. The strong positive correlation indicates that trading intensity is consistent across token types—wallets that transact large volumes of Ethereum also tend to transact large volumes of stablecoins. This collinearity suggests a shared underlying behavioral type. Fig. ~\ref{fig:scatterplot_sep_ses} reveals meaningful structure in these trading patterns. Wallets in the upper-left quadrant sell large amounts of Ethereum but smaller amounts of stablecoin per transaction, while those in the lower-right do the opposite. This separation hints at distinct user segments, potentially reflecting risk tolerance, market roles, or portfolio strategies, and supports clustering of wallets by transaction behavior.

\begin{table}[]
  \centering
  \caption{Skewness of wallet-specific intercepts for conditional mean and zero-inflation parts across the four models.}
    \begin{tabular}{lcc}
    \toprule
    Model & \multicolumn{1}{l}{Conditional Mean} & \multicolumn{1}{l}{Zero-inflation} \\
    \midrule
    Ethreum Sale & -0.152 & -0.848 \\
    Stablecoin Sale & 0.177 & -0.746 \\
    Ethereum Purchase & 0.065 & -1.081 \\
    Stablecoin Purchase & 0.127 & -0.868 \\
    \bottomrule
    \end{tabular}%
  \label{tab:skew_wallet}%
\end{table}%

\subsection{Model Fit}
We will use the full model as described in \eqref{eq:glm_link_1}-\eqref{eq:glm_link_2} for our analysis. But to see the effectiveness of adding covariates and wallet-level intercepts, we also investigate two additional models. 

\begin{itemize}
    \item \emph{Model A: Time–only baseline}\\
    Zero inflation and conditional mean contains just the splines function: $\log\mu_{t}=\beta_{0}+f(t),$ $\operatorname{logit}\pi_{t}=\gamma_{0}+g(t).$
    \item\emph{Model B: Time and covariates}\\
    Adds price and macro information but no wallet heterogeneity: $ \log\mu_{t}=\beta_{0}+f(t)+\bm\zeta^{\!\top}\mathbf X_{t},$ $\operatorname{logit}\pi_{t}=\gamma_{0}+g(t)+\bm\kappa^{\!\top}\mathbf X_{t}.$
    \item\emph{Full Model}\\
    Same fixed effects as Model B plus wallet–specific intercepts  \((\alpha_i,\gamma_i)\) as defined in ~\eqref{eq:glm_link_1}–\eqref{eq:glm_link_2}.
\end{itemize}

Table~\ref{tab:model_comparision} reports the maximized log-likelihood, residual degrees of freedom, AIC and BIC for the three models. While the improvement in AIC and BIC is marginal when update Model A to Model B by adding covariates, we see an average improvement of around 11.5\% and 7.3\% in AIC and BIC when adding wallet specific intercepts. 

\begin{table}[]
  \centering
  \caption{Model summary for the three fitted models.}
    \begin{tabular}{lrrr}
    \toprule
    & \multicolumn{3}{c}{Ethereum Sale Model} \\
          & \multicolumn{1}{l}{Model A} & \multicolumn{1}{l}{Model B} & \multicolumn{1}{l}{Full Model} \\
    \midrule
    Negative Log-Likelihood & 146445.5 & 146403.1 & 128344.5 \\
    Residual Degrees of Freedom & 168337 & 168333 & 167115 \\
    AIC   & 292937.1 & 292860.1 & 259179.0 \\
    BIC   & 293167.8 & 293131.1 & 271671.1 \\
    \bottomrule
    \end{tabular}%
  \vspace{10pt}
  
    \begin{tabular}{lrrr}
          & \multicolumn{3}{c}{Ethereum Purchase Model} \\
          & \multicolumn{1}{l}{Model A} & \multicolumn{1}{l}{Model B} & \multicolumn{1}{l}{Full Model} \\
    \midrule
    Negative Log-Likelihood & 143148.0 & 143129.3 & 125451.3 \\
    Residual Degrees of Freedom & 168337 & 168333 & 167115 \\
    AIC   & 286342.0 & 286312.6 & 253392.5 \\
    BIC   & 286572.8 & 286583.5 & 265884.7 \\
    \bottomrule
    \end{tabular}%
  \label{tab:model_comparision}%
\end{table}%

For ease of presentation, we provide the coefficients of the conditional mean model for the Ethereum sales data in Table~\ref{tab:eth_s_cond_coeff} based on Model B. In Table~\ref{tab:eth_coeff}, we provide the coefficient estimates in the Ethereum models using Model B. The key point to note here is the sign of the \texttt{ETHPRICE} covariate in all the four models. For both the zero-inflation model in purchase/sale model, we note negative coefficient for the \texttt{ETHPRICE} covariate. This means that a higher spot price lowers the chance of a structural zero in transactions and raises the probability that a wallet makes a transaction (buy or sell). This is consistent with the ``attention” effect \cite{barber_odean}, when prices rise, more wallets whether motivated by profit-taking or by fear of missing out log on and submit orders.

\begin{table}[]
  \centering
  \caption{Parameter estimates for the conditional mean component in Ethereum wallet activity.}
    \begin{tabular}{lcccc}
    \toprule
          & Estimate & Std. Error & z value & p-value \\
    \midrule
    Intercept & 13.374 & 0.182 & 73.301 & $<$2E-16 \\
    Spline Knot 1 & -0.684 & 0.185 & -3.696 & 2.19E-04 \\
    Spline Knot 2 & -1.016 & 0.168 & -6.030 & 1.64E-09 \\
    Spline Knot 3 & -6.587 & 0.264 & -24.921 & $<$ 2E-16 \\
    Spline Knot 4 & -3.836 & 0.214 & -17.922 & $<$ 2E-16 \\
    Spline Knot 5 & -5.825 & 0.169 & -34.408 & $<$ 2E-16 \\
    Spline Knot 6 & -3.170 & 0.246 & -12.904 & $<$ 2E-16 \\
    Spline Knot 7 & -5.864 & 0.193 & -30.390 & $<$ 2E-16 \\
    Spline Knot 8 & -5.658 & 0.256 & -22.109 & $<$ 2E-16 \\
    Spline Knot 9 & -7.490 & 0.224 & -33.446 & $<$ 2E-16 \\
    Spline Knot 10 & -6.356 & 0.234 & -27.105 & $<$ 2E-16 \\
    \texttt{ETHPRICE} & -1.075 & 0.051 & -21.034 & $<$ 2E-16 \\
    \texttt{RF6M} & -0.256 & 0.017 & -15.508 & $<$ 2E-16 \\
    \bottomrule
    \end{tabular}%
  \label{tab:eth_s_cond_coeff}%
\end{table}%

\begin{table}[]
  \centering
  \caption{Estimate of coefficients for the two covariates in both Ethereum sale and Ethereum purchase models}
    \begin{tabular}{lrr|rr}
          & \multicolumn{2}{c|}{Zero Inflation} & \multicolumn{2}{c}{Conditional Mean} \\
          & \multicolumn{1}{l}{Sale} & \multicolumn{1}{l|}{Purchase} & \multicolumn{1}{l}{Sale} & \multicolumn{1}{l}{Purchase} \\
    \midrule
    \texttt{ETHPRICE} & -0.077 & -0.050 & -1.075 & 0.344 \\
    \texttt{RF6M} & -0.036 & -0.010 & -0.256 & -0.255 \\
    \end{tabular}%
  \label{tab:eth_coeff}%
\end{table}%

On the other hand, the sign for the conditional mean flips. A negative coefficient is the sale model implies the transaction quantity on a single day decreases as price increases, while the positive coefficient in the purchase model implies the exact opposite, as Ethereum price increases, the purchase quantities rise. Many sellers are effectively cash extractors, developers paid in Ethereum or miners liquidating block rewards. Since their USD target is generally fixed, a higher ETH/USD price means that fewer tokens are required to hit that dollar amount, so the conditional mean drops. In contrast, wallets buying ETH are typically speculators or treasury accounts adding to long positions. Rising prices confirm their bullish view and strengthen momentum expectations and thus, the conditional mean of buys grows with Ethereum price.

\begin{figure}
\centering
\begin{subfigure}{.22\textwidth}
  \centering
  \includegraphics[width=\linewidth]{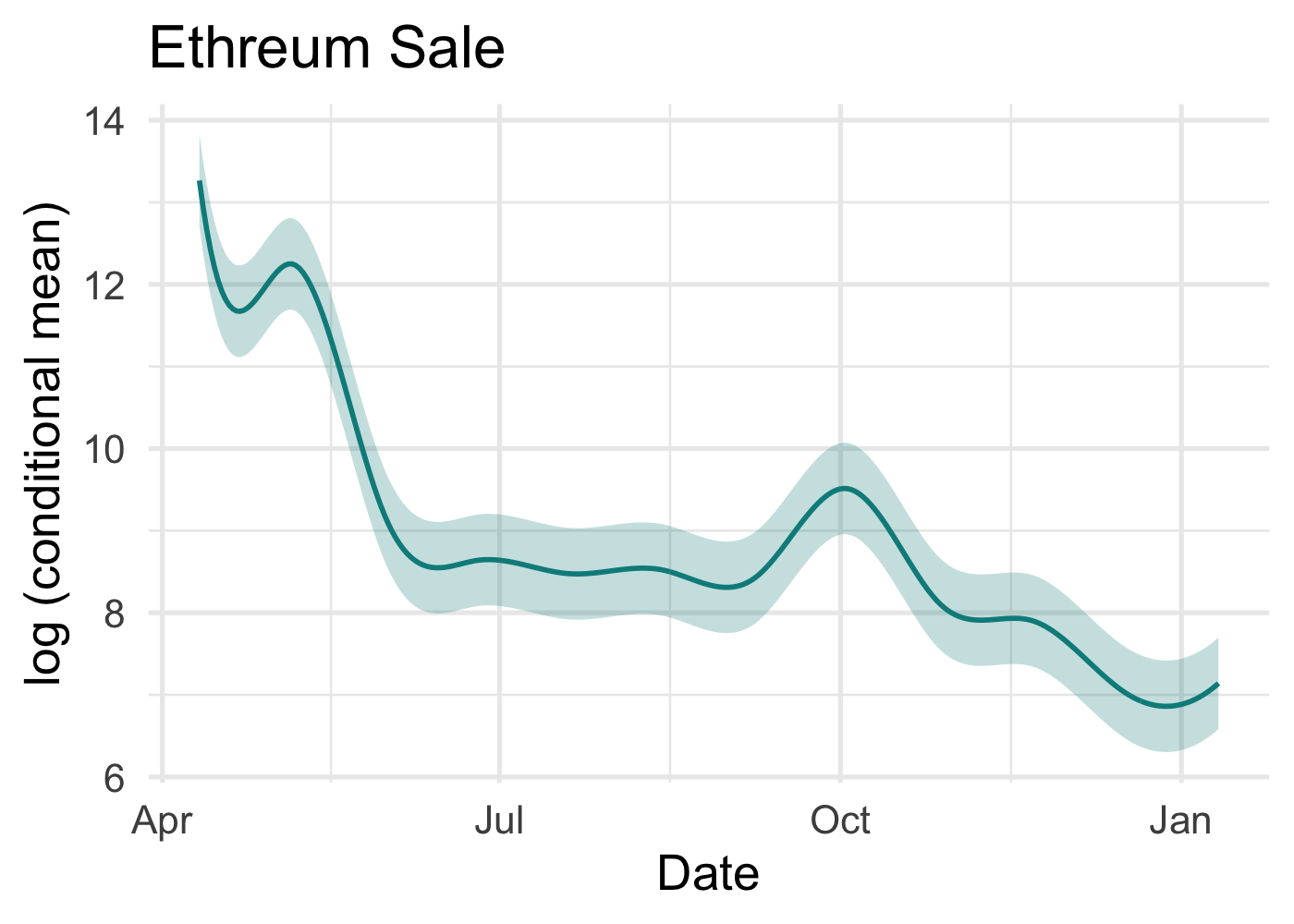}
  \caption{}
  \label{fig:es_ci}
\end{subfigure}%
\begin{subfigure}{.22\textwidth}
  \centering
  \includegraphics[width=\linewidth]{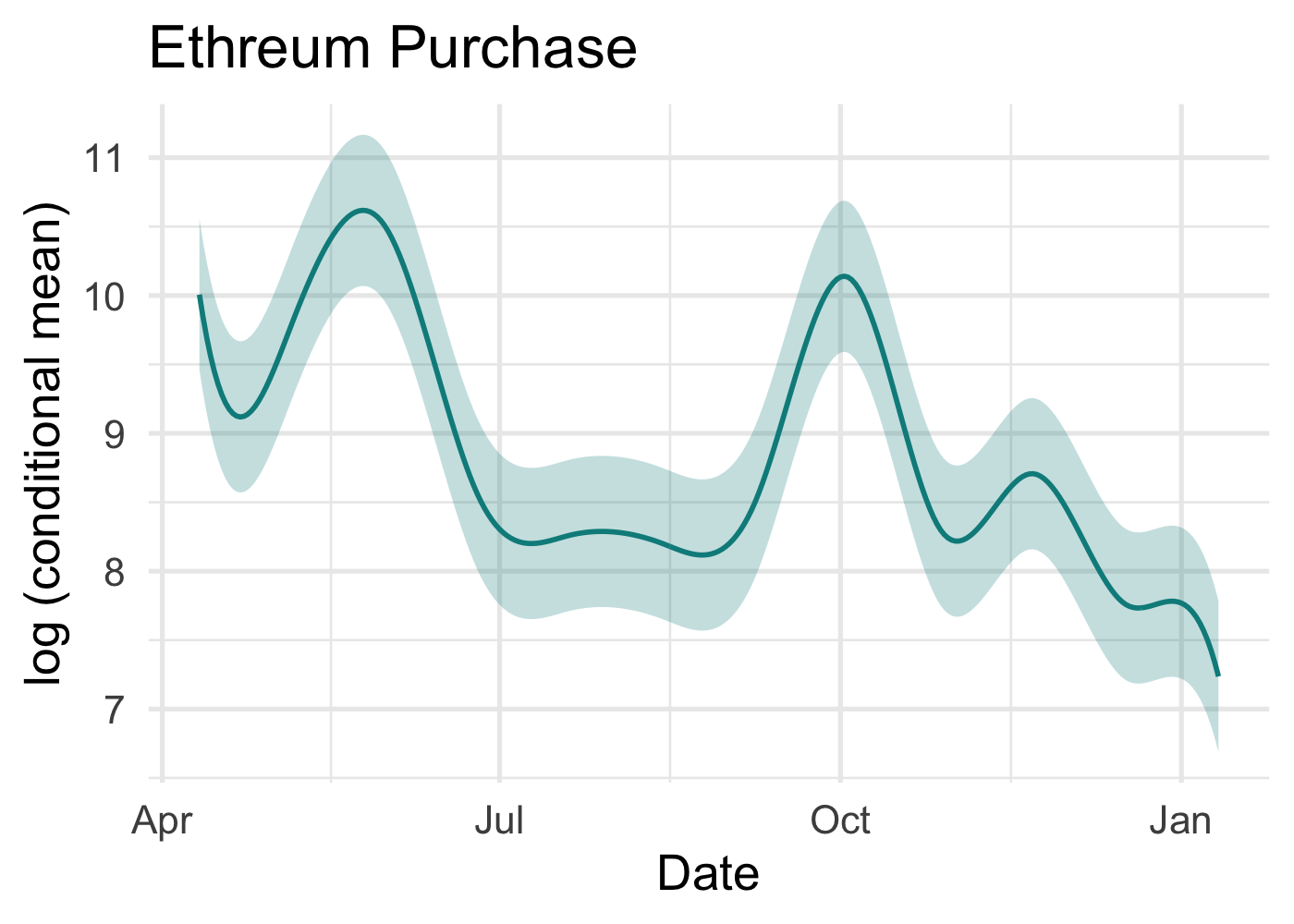}
  \caption{}
  \label{fig:ep_ci}
\end{subfigure}
\caption{Simultaneous confidence intervals for the fitted time spline functions in the conditional mean models. Panel (a) shows Ethereum sales, and panel (b) Ethereum purchases.}
\label{fig:ci_plots}
\end{figure}

Fig.~\ref{fig:ci_plots} shows the 95\,\% simultaneous confidence bands for the fitted time–spline $f(t)=\log\mu_t$ in the conditional mean component using Model B.
The bands were obtained with a parametric bootstrap ($B=1{,}000$ resamples) and calibration described in the previous section which controls the error over the entire duration of the study.

For most of the study period the confidence ribbons are very tight, signaling that the spline is estimated precisely once market covariates are taken into account. During the period of the FTX bankruptcy, however, both panels show a pronounced effect: the entire post-crisis band sits above the top of the pre-crisis band. Since the two ribbons do not overlap, at the 5\% family-wise level, the difference in conditional means is statistically significant.

\subsection{Effect of the FTX Collapse}

\begin{figure}[]
    \centering
        \begin{subfigure}{.22\textwidth}
  \centering
  \includegraphics[width=\linewidth]{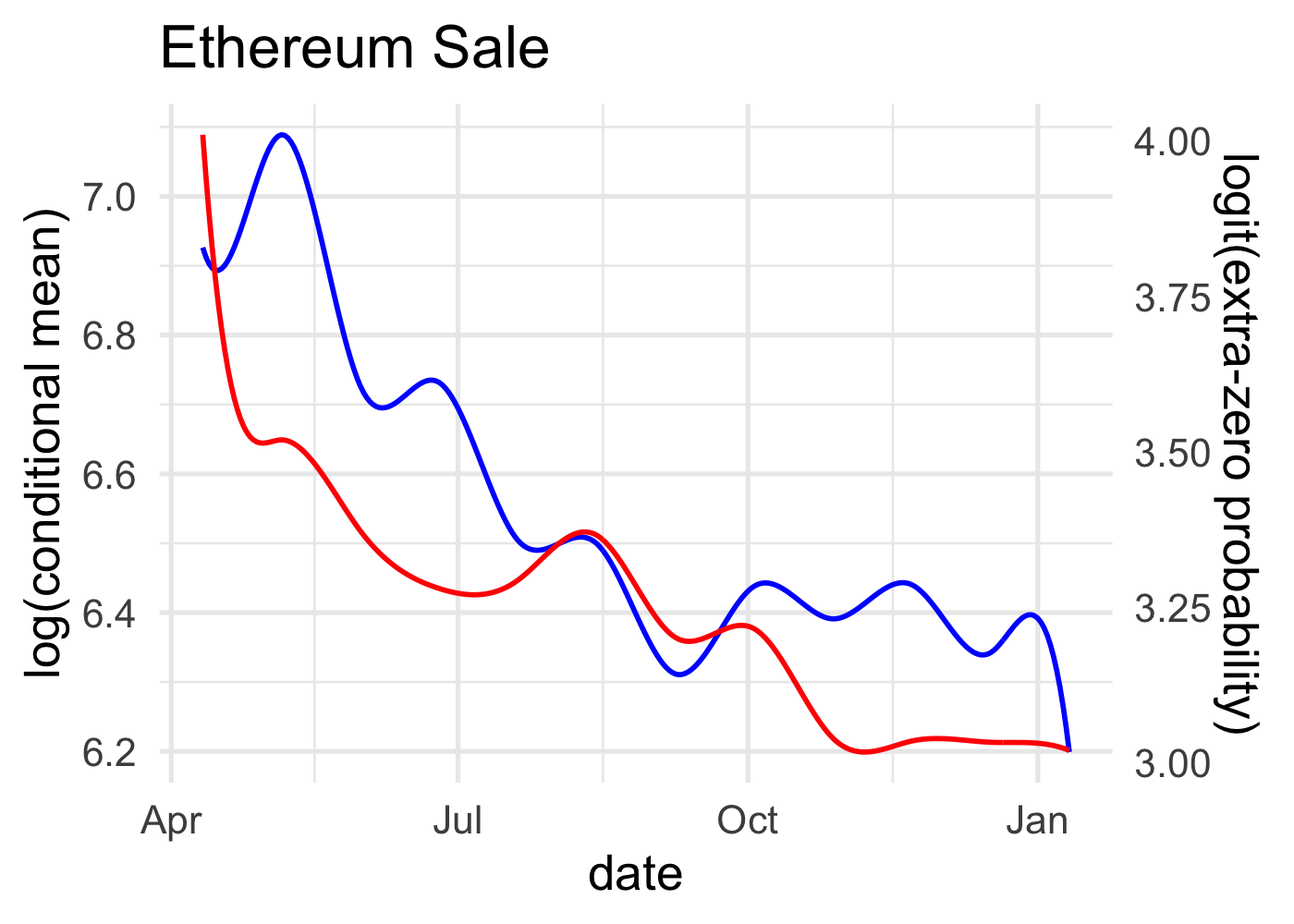}
  \caption{}
  \label{fig:sub1}
\end{subfigure}%
\begin{subfigure}{.22\textwidth}
  \centering
  \includegraphics[width=\linewidth]{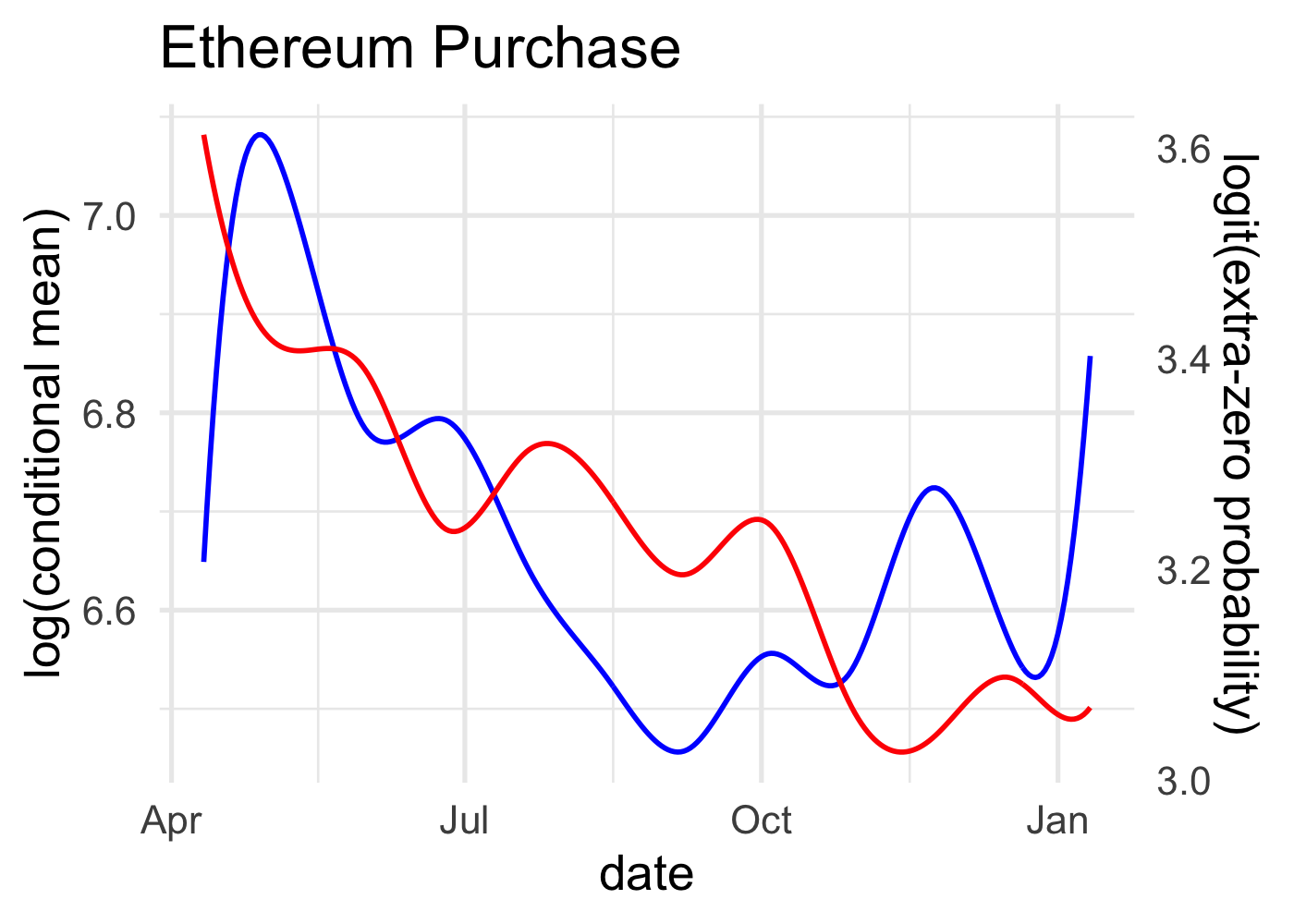}
  \caption{}
  \label{fig:sub2}
\end{subfigure}
\begin{subfigure}{.22\textwidth}
  \centering
  \includegraphics[width=\linewidth]{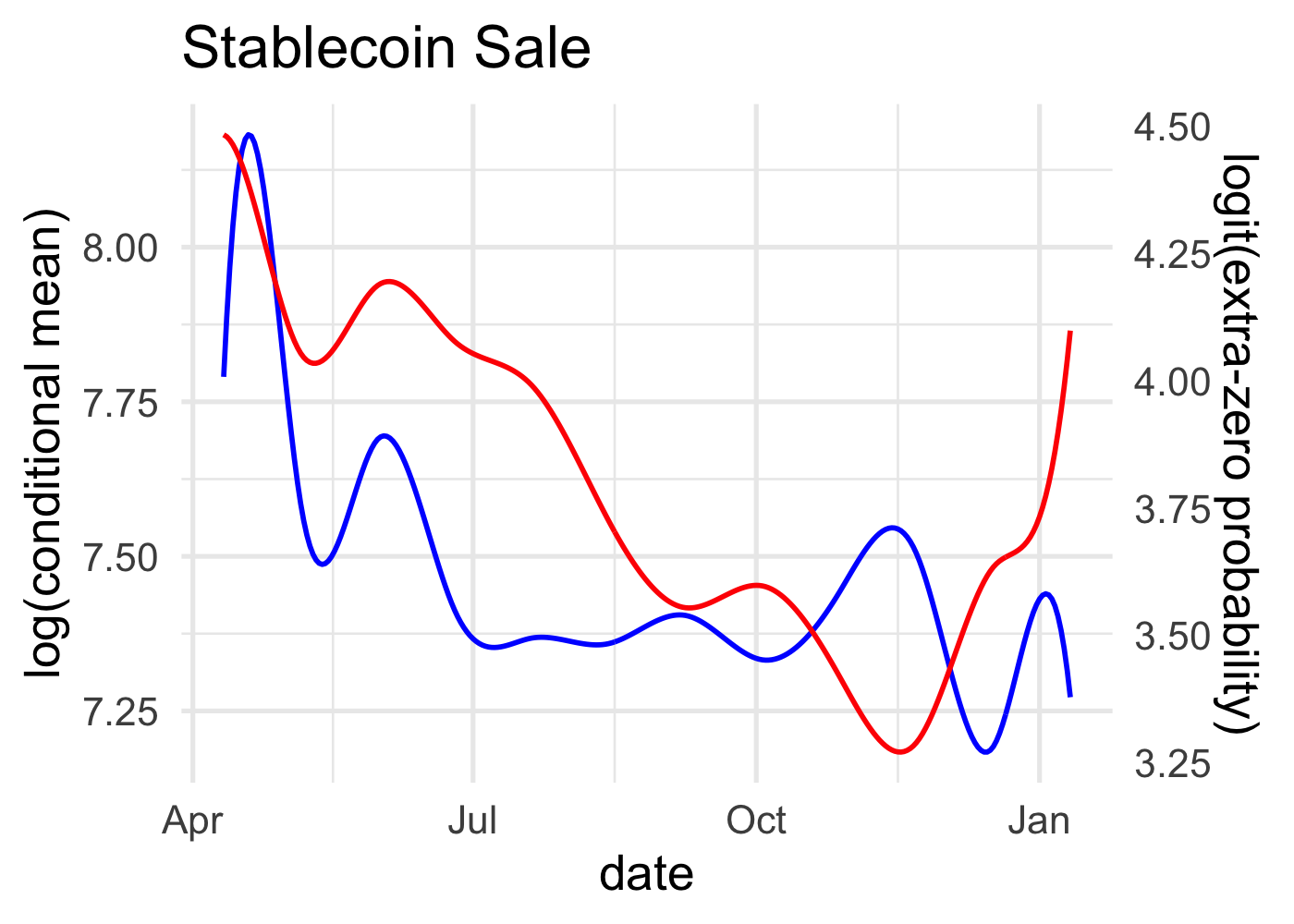}
  \caption{}
  \label{fig:sub2}
\end{subfigure}
\begin{subfigure}{.22\textwidth}
  \centering
  \includegraphics[width=\linewidth]{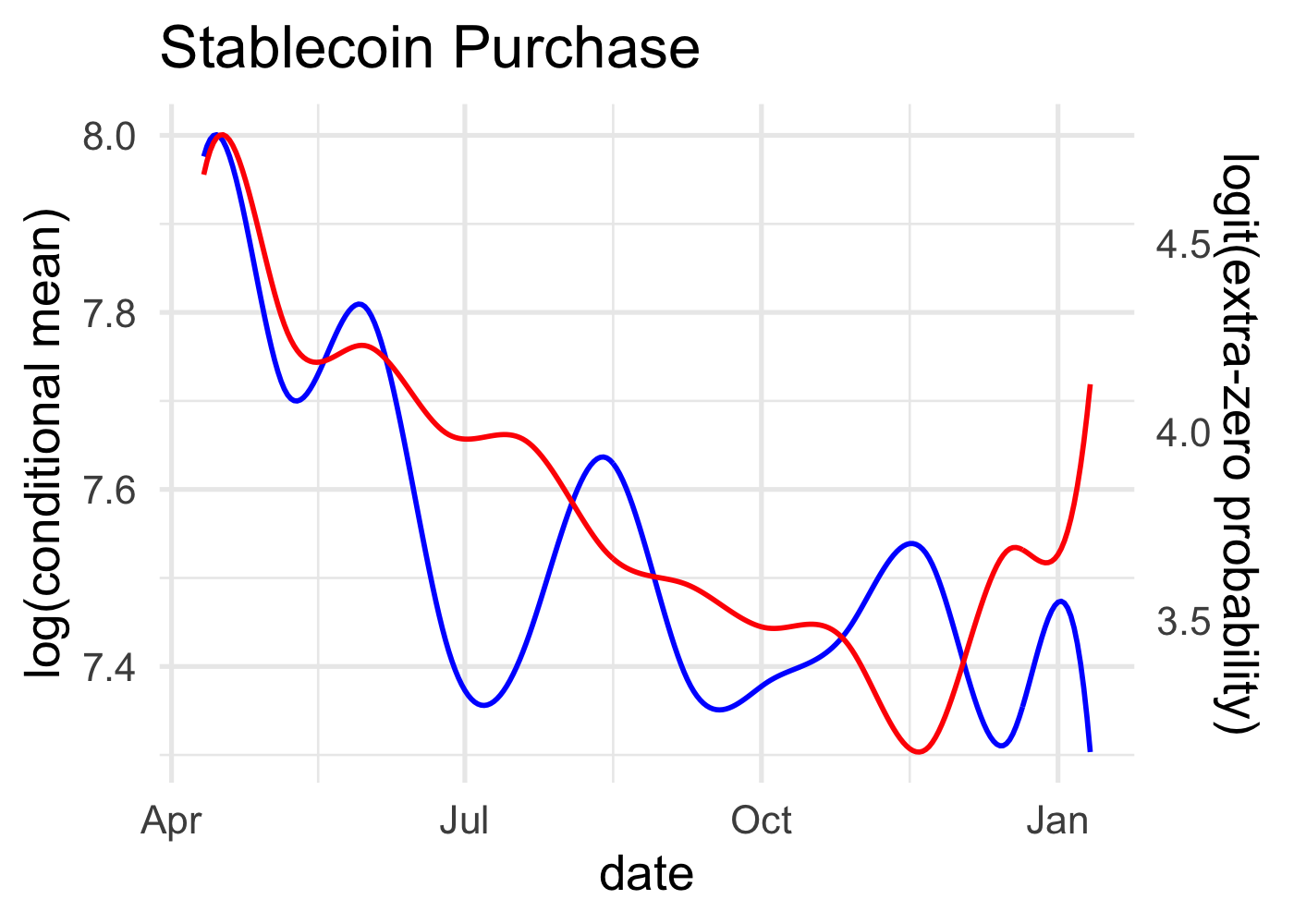}
  \caption{}
  \label{fig:sub2}
\end{subfigure}
    \caption{Transaction activity model: splines function estimate, zero-inflation model $\operatorname{logit}\pi_t$ (in red), conditional mean model $\log \mu_t$ (in blue) (a) Ethereum sale, (b) Ethereum purchase, (c) Stablecoin sale, (d) Stablecoin purchase}
    \label{fig:eth_bs}
\end{figure}

Fig. ~\ref{fig:eth_bs} display the fitted time-spline components for Ethereum and stablecoin transactions, respectively, split into sales (left panels) and purchases (right panels). Across the $276$ days and $610$ sampled wallets, there were only $7.21\%$ and $6.88\%$ of wallet-day observations was nonzero for ETH sale and ETH purchase. The numbers were meaningfully less at $3.70\%$ and $3.63\%$ for the stablecoin. 

Across both Ethereum and stable-coin streams we observe a sharp rise in trading activity in the weeks immediately surrounding the FTX collapse. For ETH sales, after the initial spike, marked by a sudden drop of the red curve in Fig.~\ref{fig:eth_bs}(a), it never returns to its pre-crisis level. The higher intensity persists throughout the remainder of the sample, indicating that once wallets began selling in response to the crisis they continued to do so on a daily basis. This persistence is consistent with a prolonged portfolio re-balancing motive.
Many wallets, especially those paid in ETH or holding staking rewards treat the episode as a warning signal and continue to convert a trickle of coins into dollars even after headline subsides.

In contrast, the buy-side red curve Fig.~\ref{fig:eth_bs}(b) trends upwards after the crash. The initial surge in participation around the event is followed by a decline as shown in Fig.~\ref{fig:eth_bs}(b), suggesting that buy-side interest dried up once the immediate volatility had passed.

Stable-coin behavior shows a different picture. Trading activity in both buying and selling spikes in mid-November but then contracts sharply. Table \ref{tab:window_mean} breaks the sample into three windows: Pre-FTX Collapse (1 Sep–30 Sep 2022), FTX Collapse (16 Oct–15 Nov 2022), and Post-FTX Collapse (1 Dec–31 Dec 2022), which corroborates the visual trends observed in the figures. Conditional mean (blue curve) for stable-coin trades peaks during the event window and falls by roughly 9–11\% afterwards. Trading propensity  drops from about 3.3\% of wallets per day in the event window to 2.6\% post-crisis.

\begin{table}[]
  \centering
  \caption{Probability of transaction and the conditional mean for the four models — Ethereum Sale, Ethereum Purchase, Stablecoin Sale, and Stablecoin Purchase over three periods: Pre-FTX Collapse (1 Sep 2022–30 Sep 2022), FTX Collapse (16 Oct 2022–15 Nov 2022), and Post-FTX Collapse (1 Dec 2022–31 Dec 2022).}
    \begin{tabular}{lcc}
    \toprule
    \multicolumn{1}{c}{Window} & Probability of Transaction & Conditional Mean \\
    \midrule
    \multicolumn{3}{c}{Ethereum Sale} \\
    \midrule
    Pre-FTX Collapse & 3.88\% & 572.8 \\
    FTX Collapse & 4.54\% & 607.1 \\
    Post-FTX Collapse & 4.59\% & 582.3 \\
    \midrule
    \multicolumn{3}{c}{Ethereum Purchase} \\
    \midrule
    Pre-FTX Collapse & 3.85\% & 660.0 \\
    FTX Collapse & 4.42\% & 716.9 \\
    Post-FTX Collapse & 4.38\% & 730.4 \\
    \midrule
    \multicolumn{3}{c}{Stablecoin Sale} \\
    \midrule
    Pre-FTX Collapse & 2.73\% & 1603.5 \\
    FTX Collapse & 3.25\% & 1743.6 \\
    Post-FTX Collapse & 2.68\% & 1439.8 \\
    \midrule
    \multicolumn{3}{c}{Stablecoin Purchase} \\
    \midrule
    Pre-FTX Collapse & 2.78\% & 1597.8 \\
    FTX Collapse & 3.34\% & 1746.9 \\
    Post-FTX Collapse & 2.66\% & 1589.9 \\
    \bottomrule
    \end{tabular}%
  \label{tab:window_mean}%
\end{table}%

In other words, wallets turned to stablecoins for immediate liquidity during the height of the turmoil, but once the market stabilized they reduced both the frequency and the size of stablecoin transactions. Ethereum sales, by contrast, maintained the elevated pace, consistent with continued portfolio rebalancing away from risk assets even after the initial shock had passed. By decomposing activity into a trade/no-trade decision and a conditional size, the zero-inflated model captures both aspects of the trading behavior, and shows that the lasting impact of the FTX event falls much more on Ethereum than dollar-pegged tokens.

\section{Conclusion and Future Work} 

In this paper, we examined how Ethereum wallet behavior changed following the collapse of FTX, a significant event that disrupted the crypto ecosystem in late 2022. Using transaction-level data, we employed zero-inflated generalized linear models to estimate both average and wallet-specific responses. Our framework accounts for sparse activity, wallet-level heterogeneity, and macroeconomic covariates such as Ethereum price and the six-month U.S. Treasury yield, allowing us to separate wallet-specific behavior from broader economic effects.

We find that wallet activity declined sharply after the collapse. The share of active wallets fell from 13 percent to 9 percent, and transaction counts dropped in two phases: a steep decline immediately following the event, followed by a slower contraction. While these aggregate trends are informative, our analysis also reveals considerable variation across users. Some wallets reduced activity significantly or exited the network, while others maintained or increased engagement. These findings highlight how aggregate or price-based metrics can obscure meaningful wallet-level behavior, particularly for understanding shocks to the Ethereum ecosystem.

Looking ahead, extending this analysis across a longer time horizon would help assess whether and how wallet activity recovers. A multi-year panel could enable the study of churn, recovery, and longer-term behavioral shifts. Future work will focus on modeling the joint dynamics of multiple wallet behaviors within a unified framework. Simultaneous modeling of these behaviors helps capture cross-transaction dependencies and better identify joint patterns in response to shocks. In addition, incorporating hazard models could help quantify wallet exit rates and the timing of inactivity following major disruptions.

By emphasizing interpretability and heterogeneity, this work contributes to explainable AI approaches for blockchain systems. Our results offer practical insight for researchers, regulators, and infrastructure developers seeking to understand decentralized user behavior during periods of stress. Wallet-level analysis offers a valuable perspective for evaluating how users respond to major market events.

\vspace{12pt}
\color{red}


\begin{thebibliography}{00}


\bibitem{coinbase_state_of_crypto_2025}
Coinbase. ``The state of crypto summit 2025: key insights on defining a new global financial system''. \textit{Coinbase Blog}, June 23, 2025. \url{https://www.coinbase.com/blog/state-of-crypto-2025-summary}.

\bibitem{cucuringu_clustering_uniswap_traders} Miori, D., Cucuringu, M. ``Clustering Uniswap v3 traders from their activity on multiple liquidity pools, via novel graph embeddings''. Digital Finance 6, 113–143 (2024). 

\bibitem{bolker2009glmm} Bolker, B. M., et al. ``Generalized linear mixed models: a practical guide for ecology and evolution''. Trends in Ecology \& Evolution, 2009.


\bibitem{FTX_downfall} D. Vidal-Tomás, A. Briola, and T. Aste, ``FTX’s downfall and Binance’s consolidation: The fragility of centralised digital finance''. Physica A: Statistical Mechanics and its Applications, vol. 625, p. 129044, 2023.

\bibitem{crypto_volume_volatility} P. K. Mishra and S. K. Das, ``Unravelling the volume-volatility nexus in cryptos under structural breaks using fat-tailed distributions: mixture of distribution hypothesis and implications for market efficiency''. Emerging Markets Finance and Trade, vol. 61, no. 3, pp. 712–731, 2025.

\bibitem{crypto_volatility} A. Suto and T. Yamamoto, ``Financial Markets Effect on Cryptocurrency Volatility: Pre- and Post-Future Exchanges Collapse Period in USA and Japan''. Journal of Risk and Financial Management, vol. 13, no. 1, p. 24, 2025.

\bibitem{bayesian_structural_counterfactual} Khan, K., Khurshid, A. and Cifuentes-Faura, J. ``Causal estimation of FTX collapse on cryptocurrency: a counterfactual prediction analysis''. Financ Innov 2025.

\bibitem{xblock_eth} P. Zheng, Z. Zheng, J. Wu, and H.-N. Dai, ``XBlock-ETH: Extracting and Exploring Blockchain Data From Ethereum''. IEEE Open Journal of the Computer Society, vol. 1, pp. 95–106, 2020.

\bibitem{barber_odean} B. M. Barber and T. Odean, ``All that glitters: the effect of attention and news on the buying behavior of individual and institutional investors,” The Review of Financial Studies, vol. 21, no. 2, pp. 785–818, Apr. 2008.

\bibitem{mcgillycuddy2025parsimoniously}
M. McGillycuddy, G. Popovic, B. M. Bolker, and D. I. Warton, ``Parsimoniously fitting large multivariate random effects in glmmTMB''. Journal of Statistical Software, vol. 112, pp. 1–19, 2025.





\end{thebibliography}
\end{document}